\documentclass[paper,prd,twocolumn,nofootinbib,preprintnumbers,amsmath,amssymb]{revtex4-2}

\usepackage{latexsym,graphicx,amssymb,amsmath,mathrsfs,bbold} 
\usepackage{slashed,mathbbol}
\usepackage[utf8]{inputenc}
\usepackage{color}
\usepackage{makecell}

\newcommand{\be}{\begin{equation}}      
\newcommand{\ee}{\end{equation}}      
\newcommand{\bea}{\begin{eqnarray}}      
\newcommand{\eea}{\end{eqnarray}}    
\newcommand{\Tr}{\,\textrm{Tr}\,}
\newcommand{\Log}{\,\textrm{Log}\,}
\newcommand{\tr}{\,\textrm{tr}\,}
\renewcommand{\log}{\,\textrm{log}\,}
\newcommand{\un}{\mbox{$\mathbb{1}$}}
\newcommand{\ch}{\,\textrm{ch}\,}
\newcommand{\an}{\,\textrm{an}\,}
\renewcommand{\un}{\mbox{$\mathbb{1}$}}
\renewcommand{\iff}{\,\textrm{if}\,}  
\newcommand{\elsee}{\,\textrm{else}\,}  
\newcommand{\MeV}{\,\textrm{MeV}\,}  
\newcommand{\crit}{\,\textrm{crit}\,}

\makeatletter
\renewcommand\appendix{\par
\setcounter{section}{0}%
\setcounter{subsection}{0}%
\gdef\thesection{\appendixname\space\@Alph\c@section}}

\long\def\unmarkedfootnote#1{{\long\def\@makefntext##1{##1}\footnotetext{#1}}}
\makeatother

\begin{document} 

\title{Order of the $SU(N_f)\times SU(N_f)$ chiral transition via\\ the functional renormalization group} 
\author{G. Fej\H{o}s}
\email{gergely.fejos@ttk.elte.hu}
\affiliation{Institute of Physics and Astronomy, E\"otv\"os University, 1117 Budapest, Hungary,}
\affiliation{Interdisciplinary Theoretical and Mathematical Sciences Program (iTHEMS), RIKEN, Wako, Saitama 351-0198, Japan}
\author{T. Hatsuda}
\email{thatsuda@riken.jp\newline}
\affiliation{Interdisciplinary Theoretical and Mathematical Sciences Program (iTHEMS), RIKEN, Wako, Saitama 351-0198, Japan}

\begin{abstract}
Renormalization group flows of the $SU(N_f)\times SU(N_f)$ symmetric Ginzburg-Landau potential are calculated for a general number of flavors, $N_f$. Our approach does not rely on the $\epsilon$ expansion, but uses the functional renormalization group, formulated directly in $d=3$ spatial dimensions, with the inclusion of all possible (perturbatively) relevant and marginal operators, whose number is considerably larger than those in $d=4$. We find new, potentially infrared stable fixed points spanned throughout the entire $N_f$ range. By conjecturing that the thermal chiral transition is governed by these ``flavor continuous" fixed points, stability analyses show that for $N_f\geq 5$ the chiral transition is of second order, while for $N_f=2,3,4$, it is of first order. We argue that the $U_{\rm A}(1)$ anomaly controls the strength of the first order chiral transition for $N_f=2,3,4$, and makes it almost indistinguishable from a second order one, if it is sufficiently weak at the critical point. This could open up a new strategy to investigate the strength of the $U_{\rm A}(1)$ symmetry breaking around the critical temperature.
\end{abstract}

\maketitle

\section{Introduction}

Since the seminal paper of Pisarski and Wilczek \cite{pisarski84} it has been widely accepted that the chiral phase transition of quantum chromodynamics (QCD) is of fluctuation-induced first order in the chiral limit for $N_f > 2$ quark flavors. For $N_f=2$, the transition order depends on the strength of the $U_{\rm A}(1)$ anomaly at the critical point in the underlying theory before the dimensional reduction: Only if $m_{\eta'}(T_c) \gg T_c$ does the Ginzburg-Landau potential acquire $O(4)$ symmetry and predict the transition to be of second order. The original argument of \cite{pisarski84} was based on the absence of infrared stable fixed points in the $\epsilon$ expansion around $d=4$ of the renormalization group (RG) flows.
RG studies directly at $d=3$ either by the perturbative approach with higher-loop contributions \cite{butti03} or by the nonperturbative functional renormalization group (FRG) technique \cite{fukushima11,grahl13,fejos14,resch19,braun24} also seemed to confirm the original scenario: However, those approaches did not consider all relevant and marginal operators at $d=3$, therefore, the fixed-point structure of the full parameter space has not been fully explored.

If the chiral transition is of first order for $N_f > 2$, as predicted by the $\epsilon$ expansion, then there must exist a critical quark mass, at which the chiral transition changes from crossover ($m_{q} > m_{q,\crit}$) to first order ($m_{q} < m_{q,\crit}$). Several lattice QCD simulations attempted to determine $m_{q,\crit}$ \cite{karsch01,deforcrand03,jin15,jin17,bazazov17}, but they typically showed strong cutoff and discretization dependencies. Recently, lattice QCD simulations with unimproved staggered fermions suggested that the chiral transition could be of second order in the chiral limit, presumably for any flavor number up to the conformal window, i.e., for $N_f<N_f^*\sim 9$-$12$ \cite{cuteri21}. A study using highly improved staggered quark action did not find direct evidence of the first order transition for $N_f=3$ in the range of the pion mass $80\MeV \lesssim m_{\pi} \lesssim 140 \MeV$ \cite{dini21}. With the use of Möbius domain wall fermions, the critical quark mass for $N_f=3$ was estimated to be $m_{q,\crit} \lesssim 4 \MeV$ \cite{zhang24}. Apart from lattice studies, a recent work using the Dyson-Schwinger approach also predicted the absence of a first order transition for $N_f=3$ \cite{bernhardt23}. Furthermore, nonperturbative computations with the numerical conformal bootstrap also claimed that the transition can be of second order for $N_f=3$ \cite{kousvos22}. The increasing number of pieces of evidence of a possible second order transition for $N_f >2$, in contrast to the prediction of the $\epsilon$ expansion, is puzzling.

One of the present authors calculated the renormalization group flows of couplings of the $N_f=3$ Ginzburg-Landau potential in a truncation, where all terms up to ${\cal O}(\phi^6)$ in $d=3$ were included \cite{fejos22}. The necessity of such an approximation was based on the expectation that at least all relevant and marginal interactions around the Gaussian fixed point should be taken into account (the marginal interactions contain a number of six fields). This study was performed directly in $d=3$ dimensions, without the use of the $\epsilon$ expansion, employing the FRG technique. The main finding was that there does exist an infrared stable fixed point, which may potentially correspond to a second order chiral transition for $N_f=3$, but only if the $U_{\rm A}(1)$ anomaly {\it vanishes} at the critical point. This is in contrast to the results of \cite{pisarski84} and \cite{grahl13} for $N_f=2$, where the transition was predicted to be of second order [belonging to the $O(4)$ universality class] only if the $U_{\rm A}(1)$ anomaly is {\it strong} enough.

The thermal fate of the $U_{\rm A}(1)$ symmetry is still under debate \cite{lahiri21}.  There are studies finding that the axial anomaly is still relevant at the critical temperature \cite{dick15,ding21,kaczmarek21,bazazov12,bha14,kaczmarek23}, while some others claim that the $U_{\rm A}(1)$ symmetry is in effect restored at that point \cite{dini21,brandt16,tomiya16,aoki21,aoki22}. We also note that the thermal behavior of the anomaly could be very different for vanishing quark masses compared to the physical point \cite{fejos23}. 

Our goal in the present study is to extend the previous results of \cite{fejos22} to an arbitrary number of quark flavors. To this end, we use the FRG technique in the local potential approximation. The free energy functional will be expanded up to ${\cal O}(\phi^6)$ so that all relevant and marginal interactions (around the Gaussian fixed point) can be included. We are interested in whether new fixed point(s) with one relevant direction can be found, which were inaccessible in earlier studies and could potentially be responsible for a second order chiral transition. 

The paper is organized as follows. In Sec. II, we set up the model and the FRG method with the corresponding truncation, putting particular emphasis on the chiral invariant structure as building blocks of the potential. We analyze in detail the main differences as the flavor number increases. In Sec. III, we present the calculations of the renormalization group flows and give the $\beta$ functions explicitly for $N_f>6$. For lower values of $N_f$, the corresponding formulas are very complicated and put into Supplemental Material. Section IV is devoted to discussing the fixed point structures, together with their stability analyses. The reader finds the conclusions in Sec. V.

\section{FRG for the $SU(N_f)\times SU(N_f)$ model}

In order to decide whether a system can undergo a second order phase transition, one needs to know if the free energy, ${\cal F}$, as a functional of the mean field (or implicitly an external source), can admit scaling behavior. That is, one has to search for fixed points of the renormalization group flows of ${\cal F}$. In its functional version, the renormalization group generates the scale dependence of the free energy,  denoted from now on by ${\cal F}_k$, via the Wetterich equation \cite{wetterich93},
\bea
\label{Eq:flow}
\partial_k {\cal F}_k = \frac12 \tilde{\partial}_k \Tr \Log ({\cal F}_k''+R_k),
\eea
where ${\cal F}_k''$ is the second derivative matrix of ${\cal F}_k$ with respect to all field variables, and $k$ is the scaling variable being the wave number that separates fluctuations, which are included in ${\cal F}_k$ from those that are not. This scale separation is guaranteed by the regulator matrix, $R_k$, which, in effect freezes all modes with wave numbers lower than $k$. Note that the $\tilde{\partial}_k$ operator by definition acts only on $R_k$ and both the Tr and Log operations need to be taken in the functional and matrix senses.

In the system of our interest ${\cal F}_k$ is a functional of a $N_f\times N_f$ complex matrix field $\Phi$, which emerges from the quark $\bar{q} q$ condensate of the underlying microscopic theory of quantum chromodynamics. According to the Ginzburg-Landau paradigm, close to a second order (or weakly first order) transition, at a suitable UV scale $\Lambda$, also serving as the starting point of the RG, ${\cal F}_{\Lambda}$ can be expanded in terms of the components of $\Phi$. To this end, it is convenient to use the $U(N_f)$ generators as a basis, i.e., $\Phi = \phi_a T_a \equiv (s_a+i\pi_a)T_a$, where $\Tr(T_a T_b)=\delta_{ab}/2$ [see details of the $U(N)$ algebra in Appendix A].

In this study we employ the leading order of the derivative expansion without wave function renormalization, sometimes called the local potential approximation (LPA). Omitting the wave function renormalization is equivalent of neglecting the anomalous dimension, which, based on studies on scalar models, is expected to be small\footnote{In case of $O(N)$ scalar theories, the anomalous dimension is of the order of ${\cal O}(10^{-2})$.}. That is, the scale dependent free energy is approximated as
\bea
\label{Eq:ansatz}
{\cal F}_k = \int d^3x \Big[ \Tr [\partial_i \Phi^\dagger \partial_i \Phi] + {\cal V}_k(\Phi)\Big],
\eea
where the local function ${\cal V}_k$ is called the potential, giving the free energy density for homogeneous field configurations. From here onwards, we restrict our discussion to an approximation of ${\cal V}_k$ that only contains perturbatively relevant and marginal interactions, i.e.,  in $d=3$ spatial dimensions every possible term up to the order ${\cal O}(\phi^6)$ is kept, while the remaining ones are dropped as they are irrelevant (if the anomalous dimension is small, then scaling is equivalent to that of around the Gaussian fixed point). Note that this creates space for a much larger set of interactions compared to the $\epsilon$ expansion, which operates close to $d=4$, allowing only terms up to ${\cal O}(\phi^4)$.

Since we are not to rely on the $\epsilon$ expansion, that is, all renormalization group flows will be evaluated directly in $d=3$, there is no small parameter in our approach and thus the optimization of (\ref{Eq:flow}) important. First we note that scale separation in (\ref{Eq:flow}) is achieved through the regulator term
\bea
\int_x\int_y R_k(\vec{x},\vec{y}) \Tr[\Phi^\dagger(\vec{x})\Phi(\vec{y})],
\eea
which is chirally symmetric [note that $R_k(\vec{x},\vec{y})=R_k(\vec{y},\vec{x})]$, therefore, chiral symmetry is respected throughout the RG flow. Second, it has been known for a long time that for the approximate form of (\ref{Eq:ansatz}), the $R_k(\vec{q},\vec{p}) = (2\pi)^3(k^2-\vec{q}^2)\Theta(k^2-\vec{q}^2)\delta(\vec{q}+\vec{p})$ function, defined in Fourier space, optimizes the renormalization group flow equation (\ref{Eq:flow}) \cite{litim01}.
 It is also known that this choice guarantees that the derivative expansion is converging \cite{balog19}. Assuming homogeneous field configurations for $\Phi$, this choice of regularization leads (\ref{Eq:flow}) to
\bea
\label{Eq:flow2}
k\partial_k {\cal V}_k = \frac{k^4}{12\pi^2} \tilde{\partial}_k \tr \log (k^2 + {\cal V}_k''),
\eea
where ${\cal V}_k''$ is the second derivative matrix of ${\cal V}_k$, and now the trace and log operations need to be taken in the matrix sense only. In accordance with the definition of $\tilde{\partial}_k$, it acts only on the explicit $k$ dependence, and not that of ${\cal V}_k''$.

\subsection{Basic invariants}
Let $U_{\rm L}$ and $U_{\rm R}$ be independent $N_f\times N_f$ unitary matrices. Then, a chiral transformation acts on the field as $\Phi \rightarrow U_{\rm L} \Phi U_{\rm R}^\dagger$. If we are to construct an ${\cal F}_k$ functional that respects chiral symmetry at all scales, then the potential ${\cal V}_k$ can only depend on the following combinations of $\Phi$:
\begin{subequations}
\label{Eq:inv}
\bea
I_1 &=& \Tr (\Phi^\dagger \Phi),\\
\tilde{I}_2 &=& \Tr (\Phi^\dagger \Phi\Phi^\dagger \Phi),\\
\tilde{I}_3 &=& \Tr (\Phi^\dagger \Phi\Phi^\dagger \Phi\Phi^\dagger \Phi),
\eea
\end{subequations}
where we also used our assumption that beyond ${\cal O}(\phi^6)$ all terms are to be dropped. Note that, for $N_f=2$, $\tilde{I}_3$ is not independent from $I_1$ and $\tilde{I}_2$, therefore, it must also be left out from ${\cal V}_k$. For simplicity, we will work with a modified set of invariants,
\begin{subequations}
\bea
I_1 &=& \Tr (\Phi^\dagger \Phi),\\
I_2 &=& \Tr (\Phi^\dagger \Phi-\Tr(\Phi^\dagger \Phi)/N_f)^2,\\
I_3 &=& \Tr (\Phi^\dagger \Phi-\Tr(\Phi^\dagger \Phi)/N_f)^3,
\eea
\end{subequations}
which is completely equivalent to (\ref{Eq:inv}). This will turn out to be a more convenient choice as in the background $\Phi \sim \un_{N_f\times N_f}$, $\Phi^\dagger \Phi = \Tr (\Phi^\dagger \Phi)/N_f$, thus $I_2=0=I_3$.

We also need to implement the $U_{\rm A}(1)$ anomaly into the free energy. This can be achieved via the usual Kobayashi-Maskawa-`t Hooft (KMT) determinant term \cite{kobayashi70,thooft76}, i.e.,
\bea
I_{\det} = \det \Phi^\dagger + \det \Phi.
\eea
Note that $\tilde{I}_{\det} = \det \Phi^\dagger - \det \Phi$ has the wrong parity, and $\tilde{I}_{\det}^2$ is not independent from $I_{\det}$ and the previous invariants for any flavor number $N_f$. That is to say, the only way to include the anomaly is through the powers of $I_{\det}$, presumably multiplied by invariants that are chirally symmetric (i.e., $I_1$, $I_2$, $I_3$).

\subsection{Structure of the local potential}

We divide the potential into two parts,
\bea
\label{Eq:Vdiv}
{\cal V}_k = {\cal V}_{\ch,k} + {\cal V}_{\an,k},
\eea
where ${\cal V}_{\ch,k}$ respects any $U_{\rm L}(N_f)\times U_{\rm R}(N_f)$ transformation, while ${\cal V}_{\an,k}$ breaks $U_{\rm A}(1)$. We stress that the form of ${\cal V}_{\ch,k}$ is completely independent of the actual flavor number (apart from $N_f=2$, where $I_3$ needs to be left out as it is not an independent invariant), while the structure of ${\cal V}_{\an,k}$ depends on the size of $\Phi$, determined by $N_f$. By leaving out the perturbatively irrelevant interactions, 
we have
\bea
\label{Eq:Vch}
\!\!{\cal V}_{\ch,k} &=& m^2 I_1 + \nonumber\\
&+&g_1 I_1^2 + g_2 I_2 \nonumber\\
&+& \lambda_1 I_1^3 + \lambda_2 I_1 I_2 + g_3 I_3,
\eea
where in each line we have ${\cal O}(\phi^2)$, ${\cal O}(\phi^4)$, ${\cal O}(\phi^6)$ interactions, respectively. Note that, close to $d=4$ spatial dimensions, $m^2$, $g_1$, and $g_2$ are the only couplings that are not irrelevant, however, for $d=3$ there are three additional ones, i.e., $\lambda_1$, $\lambda_2$, $g_3$ (for $N_f=2$ we have to set formally $g_3\equiv 0$). Now we investigate the structure of ${\cal V}_{\an,k}$ as a function of $N_f$. Our task is to find all anomalous interactions, whose field content does not exceed ${\cal O}(\phi^6)$.

\underline{$N_f=2$}: 
In this case the $\Phi$ matrix is $2\times 2$, thus $I_{\det}$ is ${\cal O}(\phi^2)$. This yields various new interactions to emerge from the $U_{\rm A}(1)$ anomaly. Keeping only the (perturbatively) relevant and marginal terms we have
\bea
\label{Eq:Vnoan}
{\cal V}_{\an,k} &=& a I_{\det} \nonumber\\
&+&a_2 I_{\det}^2+b_1 I_1 I_{\det} \nonumber\\
&+&a_3 I_{\det}^3 + b_2 I_1^2 I_{\det} + b_3 I_1 I_{\det}^2+b_4 I_2 I_{\det},
\eea
where the terms in each line are ${\cal O}(\phi^2)$, ${\cal O}(\phi^4)$, and ${\cal O}(\phi^6)$, respectively. For $N_f=2$, the number of the relevant and marginal couplings is 12 (note again that $g_3\equiv 0$ for $N_f=2$).

\underline{$N_f=3$}: 
The $\Phi$ matrix is $3\times 3$, and as a consequence $I_{\det}$ is ${\cal O}(\phi^3)$. Dropping the irrelevant interactions we have
\bea
{\cal V}_{\an,k} = a I_{\det}+ b I_1 I_{\det}+a_2 I_{\det}^2,
\eea
each term being ${\cal O}(\phi^3)$, ${\cal O}(\phi^5)$, ${\cal O}(\phi^6)$, respectively. For $N_f=3$, the number of the relevant and marginal couplings is 9.

\underline{$N_f=4$}: 
The structure of ${\cal V}_{\an,k}$ is getting simpler as we increase $N_f$. Now there are only two anomalous combinations that are not irrelevant:
\bea
\label{Eq:anNf4}
{\cal V}_{\an,k} &=& a I_{\det} + b I_1 I_{\det}.
\eea
For $N_f=4$, the number of the relevant and marginal couplings is 8.

\underline{$N_f=5,6$}: 
The structure of ${\cal V}_{\an,k}$ is the same for both cases, as the only anomalous combination that is not irrelevant is simply the usual KMT term:
\bea
{\cal V}_{\an,k} &=& a I_{\det}.
\eea
For $N_f=5,6$, the number of the sum of the relevant and marginal couplings is 7.

\underline{$N_f>6$}: 
Finally, for large enough $N_f$, there is no way to construct any anomalous term that is not irrelevant. As a result,
\bea
{\cal V}_{\an,k} &=& 0.
\eea
For $N_f>6$, the number of the relevant and marginal couplings is 6.

In what follows, we need to project the flow equation (\ref{Eq:flow2}) onto the combination of invariants, which then leads to the logarithmic scale derivative ($k\partial_k$) of each coupling.  The $\beta$ functions are, in turn, defined as the logarithmic scale derivative of the {\it dimensionless} couplings (denoted by a bar on top), rescaled by appropriate powers of $k$. For the anomaly free part, these are
\bea
\bar{m}_k^2 &=& m_k^2/k^2, \quad \bar{g}_{1,k}=g_{1,k}/k, \quad \bar{g}_{2,k}=g_{2,k}/k,\nonumber\\
\bar{\lambda}_{1,k} &=& \lambda_{1,k}, \quad \bar{\lambda}_{2,k}=\lambda_{2,k}, \quad \bar{g}_{3,k} = g_{3,k}.
\eea
As for the anomalous part, the rescalings depend on $N_f$:
\bea
N_f &=& 2 \quad\!\! \leftarrow \!\!\quad \begin{cases}
\bar{a} = a/k^2, \quad\bar{a}_2 = a_2/k, \quad\bar{b}_1 = b_1/k, \nonumber\\
\bar{a}_3 = a_3, \quad\bar{b}_2 = b_2, \quad\bar{b}_3 = b_3, \quad\bar{b}_4 = b_4,
\end{cases}\\
N_f &=& 3 \quad\!\! \leftarrow \!\!\quad \begin{cases}
\bar{a} = a/k^{3/2}, \quad\bar{b} = b/k^{1/2}, \quad\bar{a}_2= a_2, \nonumber
\end{cases}\\
N_f &=& 4 \quad\!\! \leftarrow \!\!\quad \begin{cases}
\bar{a} = a/k, \quad\bar{b} = b,\nonumber
\end{cases}\\
N_f &=& 5 \quad\!\! \leftarrow \!\!\quad \begin{cases}
\bar{a} = a/k^{1/2}, \nonumber
\end{cases}\\
N_f &=& 6 \quad\!\! \leftarrow \!\!\quad \begin{cases}
\bar{a} = a. \nonumber
\end{cases}
\eea

\section{Coupling flows}

When we allow the $k\partial_k$ operator to act on the left-hand side (lhs) of (\ref{Eq:flow2}), we straightforwardly obtain the sum of the logarithmic scale derivative of all couplings, each multiplied by their respective invariant combination. The right-hand side (rhs) of (\ref{Eq:flow2}) should exhibit compatibility with this structure, meaning it must be a linear combination of chirally symmetric interactions in the same fashion as they appear in the lhs. Therefore, by expanding the rhs around zero field and equating the two sides, the scale dependence of the couplings should be identifiable as the coefficient of the invariant combinations in the rhs. The main technical challenge here is that the rhs of (\ref{Eq:flow2}) is {\it not manifestly} chirally invariant. This is unsurprising, as in essence, we need to evaluate the {\it tr log} of the inverse propagator matrix, and naturally the masses of the eigenmodes are not chirally invariant. However, upon performing the {\it tr log} operation, they should combine into invariants, yielding the $\beta$ functions (after also applying the appropriate rescalings of the couplings with the RG variable $k$).

Calculating the {\it tr log} of the inverse two point function, i.e., $(k^2+{\cal V}_k'')$, in the most general background field of $\Phi$ is 
a formidable task (see Appendix B). As described in detail in \cite{fejos22}, it is not necessary at all, if we do not want to prove chiral invariance, but make use of it. That is, we may capitalize on the fact that at each order the rhs of (\ref{Eq:flow2}) {\it has to} combine into a linear combination of products of invariants, thus we can (at each order) use any background field at our disposal to identify them, of course if we know the actual form of the invariants in that given background. If one follows Ref. \cite{fejos22}, then the only thing one should be cautious of is that at a given order the invariant combinations should be able to be disentangled from each other; e.g., if we only allow a simple, one component background, then this is obviously not possible. If we have, for example $\Phi=s_0T_0$, then for $N_f=2$ both $I_1=s_0^2/2$ and $I_{\det}=s_0^2/2$, showing that more components need to be introduced in $\Phi$ so that $I_1$ and $I_{\det}$ can be distinguished from each other in the rhs of (\ref{Eq:flow2}).

Unfortunately, it turns out that at higher orders even by increasing the number of independent components of $\Phi$, the above procedure might fail. The reason is that at higher orders the invariants cannot in principle be separated using a step-by-step procedure that gradually eliminates them via a clever choice of sequence of backgrounds, as done in \cite{fejos22} for $N_f=3$ specifically. A reoccurring pattern is that once we try to eliminate one of the invariants, some other(s) also diminish, making impossible to read off their respective coefficients. What we would like to stress in this study is that, fortunately, it is unnecessary to identify the invariant combinations at each order, after all. It is sufficient to obtain at a given order an enough number of linearly independent algebraic equations for the logarithmic scale derivatives of the couplings. 

The way to do so is as follows. Let us use the $\Phi = s_0T_0+s_L T_L$ background, where $T_L \equiv T_{N_f^2-1}$ is the longest (``last'') diagonal generator of the algebra. Then at ${\cal O}(\phi^n)$ all invariant combinations take the form of
\bea
\label{Eq:invcond}
\sum_{i+j=n} \alpha_{ij} s_0^i s_L^j 
\eea
with suitable $\alpha_{ij}$ coefficients. Notice that we can indeed work out the $\beta$ functions without identifying the invariant combinations, if we make use of that in both sides of the flow equation (\ref{Eq:flow2}) all invariant combinations can be expressed in forms dictated by (\ref{Eq:invcond}). The task then becomes to match the coefficients of $s_0^is_L^j$ (instead of the invariants) at each side of (\ref{Eq:flow2}), the lhs now being linear combinations of logarithmic scale derivatives of the couplings. By solving the obtained coupled equations\footnote{Some of them can be linearly dependent, but never contradictory.} we get the flow of every coupling, and after rescaling them with $k$, the $\beta$ functions themselves.

\subsection{Cases of $N_f>6$}

First we apply the outlined procedure for $N_f>6$, where the anomaly does not appear; see Eq. (\ref{Eq:Vnoan}) for the corresponding potential. At ${\cal O}(\phi^2)$ there is only one invariant, at ${\cal O}(\phi^4)$ there are two, while at ${\cal O}(\phi^6)$ we see a number of three. That is, only at ${\cal O}(\phi^4)$ and ${\cal O}(\phi^6)$ do we need coupled equations for the logarithmic scale derivatives, and the background defined above (i.e. $\Phi = s_0T_0+s_L T_L$) does provide an enough number of linearly independent equations. After a straightforward but rather complicated calculation, the $\beta$ functions are found to be the following:

\begin{widetext}
\begin{subequations}
\label{Eq:beta}
\bea
\!\!\!\!\!\!\!\!\!\!\!\!\!\!\beta_{m^2} &=& -2\bar{m}_k^2 - 2\frac{\bar{g}_{1,k}N_f (N_f^2+1)+\bar{g}_{2,k}(N_f^2-1)}{3\pi^2N_f(1+\bar{m}_k^2)^2}, \\
\!\!\!\!\!\!\!\!\!\!\!\!\!\!\beta_{g_1} &=& -\bar{g}_{1,k} + 4\frac{\bar{g}_{1,k}^2N_f^2(N_f^2+4)+2\bar{g}_{1,k}\bar{g}_{2,k}N_f(N_f^2-1)+2\bar{g}_{2,k}^2(N_f^2-1)}{3\pi^2N_f^2(1+\bar{m}_k^2)^3}- \frac{3\bar{\lambda}_{1,k}N_f(N_f^2+2)+2\bar{\lambda}_{2,k}(N_f^2-1)}{3\pi^2N_f (1+\bar{m}_{k}^2)^2},\\
\!\!\!\!\!\!\!\!\!\!\!\!\!\!\beta_{g_2} &=& -\bar{g}_{2,k} +8 \frac{3\bar{g}_{1,k}\bar{g}_{2,k} N_f+\bar{g}^2_{2,k}(N_f^2-3)}{3\pi^2N_f(1+\bar{m}_k^2)^3}-\frac{3\bar{g}_{3,k}(N_f^2-4)+\bar{\lambda}_{2,k}N_f(N_f^2+4)}{3\pi^2N_f (1+\bar{m}_k^2)^2},\\
\!\!\!\!\!\!\!\!\!\!\!\!\!\!\beta_{\lambda_1} &=&4\frac{\bar{g}_{1,k}N_f^2(3\bar{\lambda}_{1,k}N_f(N_f^2+7)+2\bar{\lambda}_{2,k}(N_f^2-1))+\bar{g}_{2,k}N_f(N_f^2-1)(3N_f\bar{\lambda}_{1,k}+4\bar{\lambda}_{2,k})}{3\pi^2N_f^3(1+\bar{m}_k^2)^3}\nonumber\\
&-&4\frac{2\bar{g}_{1,k}^3N_f^3(N_f^2+13)+6\bar{g}_{1,k}^2\bar{g}_{2,k}N_f^2(N_f^2-1)+12\bar{g}_{1,k}\bar{g}_{2,k}^2N_f(N_f^2-1)+8\bar{g}_{2,k}^3(N_f^2-1)}{3\pi^2N_f^3(1+\bar{m}_k^2)^4},\\
\!\!\!\!\!\!\!\!\!\!\!\!\!\!\beta_{\lambda_2} &=&4\frac{\bar{g}_{1,k}N_f\big(\bar{\lambda}_{2,k}N_f(N_f^2+19)+3\bar{g}_{3,k}(N_f^2-4)\big)+\bar{g}_{2,k}\big(15\bar{g}_{3,k}(N_f^2-4)+N_f(18\bar{\lambda}_{1,k}N_f+\bar{\lambda}_{2,k}(5N_f^2-1))\big)}{3\pi^2N_f^2(1+\bar{m}_k^2)^3}\nonumber\\
&-&4\frac{72N_f^2\bar{g}^2_{1,k}\bar{g}_{2,k}+6\bar{g}_{1,k}\bar{g}_{2,k}^2N_f(2N_f^2+3)+\bar{g}_{2,k}^3(24N_f^2-90)}{3\pi^2N_f^2(1+\bar{m}_k^2)^4},\\
\!\!\!\!\!\!\!\!\!\!\!\!\!\!\beta_{g_3} &=& 4\frac{5N_f\bar{g}_{1,k}\bar{g}_{3,k}+4N_f\bar{g}_{2,k}\bar{\lambda}_{2,k}+(2N_f^2-17)\bar{g}_{2,k}\bar{g}_{3,k}}{\pi^2N_f(1+\bar{m}_k^2)^3}-4\frac{54\bar{g}_{1,k}\bar{g}_{2,k}^2N_f+\bar{g}_{2,k}^3(4N_f^2-54)}{3\pi^2N_f(1+\bar{m}_k^2)^4}.
\eea
\end{subequations}
\end{widetext}
All the fixed points and the corresponding stability analyses can be found in Sec. IV. 

\subsection{Cases of $N_f\leq6$}

 The $U_{\rm A}(1)$ anomaly does fit into the UV potential for $N_f\leq 6$. In case of $N_f=5,6$ it contributes solely through the usual KMT determinant, as it is ${\cal O}(\phi^5)$ and ${\cal O}(\phi^6)$, respectively. The $N_f=5$ calculation is simpler, as what we find terms at ${\cal O}(\phi^5)$ in the expansion of the rhs of (\ref{Eq:flow2}) produce solely the KMT term. As for $N_f=6$, however, the situation is more complicated, as the anomaly then contributes at ${\cal O}(\phi^6)$, and thus entangles with the non anomalous terms. 
  The $U_{\rm A}(1)$ breaking has two separate sources for $N_f=4$, see (\ref{Eq:anNf4}). The first one contributes at ${\cal O}(\phi^4)$, while the second one is at ${\cal O}(\phi^6)$, mixing once again with the fully chirally symmetric terms. 
  As for $N_f=3$, the situation again gets simpler, as the KMT determinant is cubic in the fields and thus at each ${\cal O}(\phi^3)$ and ${\cal O}(\phi^5)$ we only have a single invariant combination.

The most complicated is the $N_f=2$ case. At ${\cal O}(\phi^2)$, ${\cal O}(\phi^4)$, ${\cal O}(\phi^6)$ we have two, four and six invariant combinations [note that $g_3\equiv 0$ in (\ref{Eq:Vch}) for $N_f=2$], respectively. 
The choice of (\ref{Eq:invcond}) only allows for a number of four independent coupled equations at ${\cal O}(\phi^6)$, which is not enough to obtain all the RG flows at this order. We need to generalize (\ref{Eq:invcond}) such that one more component is introduced, e.g., $\Phi = s_0T_0+s_3T_3+i\pi_1 T_1$ (note that $T_L \equiv T_3$ for $N_f=2$), which leads at ${\cal O}(\phi^n)$ all invariant combinations become of the form of
\bea
\label{Eq:invcond2}
\sum_{i+j+k=n} \alpha_{ijk} s_0^i s_L^j \pi_1^k.
\eea
This choice is now able to provide enough independent equations at ${\cal O}(\phi^6)$ so that all $\beta$ functions can be obtained. 

All the outlined calculations are straightforward, but quite complicated and practically can only be carried out through symbolic programming. We do not go into any more detail to what is already presented, all the expressions for the $\beta$ functions as a function of $N_f$ can be found in the Supplemental Material.

\section{Fixed points}

Zeros of the $\beta$ functions correspond to renormalization group fixed points. At each fixed point we define an $\omega_{ij} = \partial \beta_i / \beta g_j$ stability matrix, where the shorthand notations $\beta_i$ and $g_j$ are used for all $\beta$ functions and couplings. Positive (negative) eigenvalues of $\omega_{ij}$ refer to irrelevant (relevant) directions. We are especially looking for fixed points with one relevant direction. The latter, if associated with the reduced temperature, yields infrared stability at the critical point. We are interested whether such fixed points, belonging to continuous thermal transitions can exist in the whole flavor number range.

We follow the strategy already outlined in \cite{papenbrock95}, that is, we eliminate all the perturbatively marginal couplings by solving analytically their respective fixed point equation, which are then plugged into the $\beta$ functions of the relevant couplings. As shown in \cite{papenbrock95}, this is (in part) produces higher loop contributions to the flows of the relevant interactions. Stability analyses are then carried out in the space of the latter couplings using the obtained (partially) resummed $\beta$ functions.

Even before going into a detailed analysis, it is expected that the Gaussian and Wilson-Fisher fixed points will definitely be found. The former is trivial, while the latter is due to the fact that if we set the $g_2$, $\lambda_2$, $g_3$, and all the anomalous couplings to zero, then the free energy becomes $O(2N_f^2)$ symmetric. Therefore, for any $N_f$ there should exist a fixed point of ${\cal F}_k$ that belongs to the $O(N)$ universality class with $N=2N_f^2$.

\begin{table}[t]
\centering
\vspace{0.2cm}
\label{Tab:largeN}
  \begin{tabular}{ c | c | c | c | c }
  \hline
    FP & $\bar{m}^2$ & $\bar{g}_1$ & $\bar{g}_2$ & RD \\ \hline\hline
    $O(\infty)$ & $-0.33333 $ & $4.38649$ & 0 & 2 \\ \hline
    $B^\infty_2$ & $0.039812 $ & $-7.32668$ & $6.05216$ & 2 \\ \hline
    $C^\infty_1$ & $-0.37351$ & $4.89039$ & $-0.54978$ & 1 \\
  \hline  
  \end{tabular}
  \caption{Nontrivial fixed points (FPs) in the $N_f\rightarrow \infty$ limit. Note the rescalings $\bar{g}_{1} \rightarrow \bar{g}_1/N_f^2$, $\bar{g}_{2} \rightarrow \bar{g}_2/N_f$. The number of relevant directions (RD) is shown in the last column.}
\end{table}

\begin{table}[]
\centering
\vspace{0.2cm}
  \begin{tabular}{ c | c | c | c | c | c }
  \hline
   ${\bf N_f}$ & FP & $\bar{m}^2$ & $\bar{g}_1$ & $\bar{g}_2$ & RD \\ \hline\hline
   ${\bf 50}$ & $O(2N_f^2)$ & $-0.33342$ & $0.0017538$ & 0 & 2 \\ \hline
   $''$ & $B^{50}_2$ & $0.040303$ & $-0.0029448$ & 0.12152 & 2 \\ \hline
   $''$ & $C^{50}_1$ & $-0.37509$ & 0.0019579 & $-0.011198$ & 1 \\ \hline
   $''$ & $\tilde{C}^{50}_1$ & $-0.33342$ & $0.0017556$ & $-0.000088291$ & 1 \\ \hline \hline
   ${\bf 20}$ & $O(2N_f^2)$ & $-0.33385$ & $0.010939$ & 0 & 2 \\ \hline
   $''$ & $B^{20}_2$ & $0.043192$ & $-0.018915$ & 0.31043 & 2 \\ \hline
   $''$ & $C^{20}_1$ & $-0.38411$ & $0.012287$ & $-0.030728$ & 1 \\ \hline
   $''$ & $\tilde{C}^{20}_1$ & $-0.33393$ & $0.011010$ & $-0.0014253$ & 1 \\ \hline \hline
   ${\bf 10}$ & $O(2N_f^2)$ & $-0.33492$ & $0.043430$ & 0 & 2 \\ \hline
   $''$ & $B^{10}_2$ & $0.059163$ & $-0.086421$ & $0.68317$ & 2 \\ \hline
   $''$ & $C^{10}_1$ & $-0.43356$ & $0.048876$ & $-0.082581$ & 1 \\ \hline 
   $''$ & $\tilde{C}^{10}_1$ & $-0.33641$ & $0.044669$ & $-0.012667$ & 1 \\ \hline \hline
   ${\bf 6}$ & $O(2N_f^2)$ & $-0.33516$ & $0.11855$ & 0 & 2 \\ \hline
   $''$ & $B^{6}_2$ & $0.40276$ & $-1.23414$ & $3.80527$ & 2 \\ \hline
   $''$ & $C^{6}_1$ & $1.09084$ & $-6.45942$ & $16.76628$ & 1 \\ \hline 
   $''$ & $\tilde{C}^{6}_1$ & $-0.34848$ & $0.12934$ & $-0.069536$ & 1 \\ \hline 
  \end{tabular}
  \caption{Nontrivial fixed points for various flavor numbers $N_f\geq 6$. Note that interactions caused by the axial anomaly are irrelevant (marginal) for $N_f>6$ ($N_f=6$). The number of relevant directions (RD) is shown in the last column. The $\tilde{C}_1^{N_f}$ and $O(2N_f^2)$ fixed points merge as $N_f\rightarrow \infty$.}
\end{table}

\begin{table}[t]
\centering
\vspace{0.2cm}
\label{Tab:N5}
  \begin{tabular}{ c | c | c | c | c | c | c }
  \hline
    ${\bf N_f}$ & FP & $\bar{m}^2$ & $\bar{g}_1$ & $\bar{g}_2$ & $\bar{a}$ & RD\\ \hline\hline
    ${\bf 5}$ & $O(2N_f^2)$ & $-0.33386 $ & $0.16871$ & $0$ & $0$ & $2$ \\ \hline
    $''$ & $\tilde{C}^5_1$ & $-0.36068$ & $0.19128$ & $-0.12675$ & $0$ & $1$ \\ \hline
   $''$ &  $A^5_3$ & $-0.17023$ & $0.14387$ & $-0.056313$ & $-2.79735$ & $3$ \\
   \hline
  \end{tabular}
  \caption{Nontrivial fixed points for $N_f=5$. The number of relevant directions (RD) is shown in the last column. }
\end{table}

\begin{table}[]
\centering
\vspace{0.2cm}
\label{Tab:N4}
  \begin{tabular}{ c | c | c | c | c | c | c }
  \hline
    ${\bf N_f}$ & FP & $\bar{m}^2$ & $\bar{g}_1$ & $\bar{g}_2$ & $\bar{a}$ & RD \\ \hline\hline
    ${\bf 4}$ & $O(2N_f^2)$ & $-0.32940 $ & $0.25800$ & $0$ & $0$ & $3$ $(2)$ \\ \hline
    $''$ & $\tilde{C}^4_2$ & $-0.38129$ & $0.31042$ & $-0.25480$ & $0$ & $2$ $(1)$ \\ \hline
   $''$ &  $A^4_2$ & $-0.34949$ & $0.63992$ & $-1.73326$ & $-3.82052$ & $2$ \\ \hline
   $''$ &  $\tilde{A}^4_2$ & $-0.40273$ & $0.21168$ & $0.17473$ & $-0.73657$ & $2$ \\
   \hline
  \end{tabular}
  \caption{Nontrivial fixed points for $N_f=4$. The number of relevant directions (RD) is shown in the last column. The numbers in the parentheses correspond to the relevant directions under the constraint that all directions corresponding to anomalous couplings disappear.}
\end{table}

\subsection{Case of $N_f = \infty$}
Before analyzing the $N_f$ dependence of the fixed point structure, it is instructive to perform a large $N_f$ expansion on the $\beta$ functions of (\ref{Eq:beta}). After making the $\bar{g}_{1,k} \rightarrow \bar{g}_{1,k}/N_f^2$,  $\bar{g}_{2,k} \rightarrow \bar{g}_{2,k}/N_f$ rescalings and eliminating the marginal couplings, we get the following fixed point equations:
\begin{subequations}
\bea
\label{Eq:largeNm2}
0&\equiv&\beta_{m_2}=-2\bar{m}_k^2-\frac{2(\bar{g}_{1,k}+\bar{g}_{2,k})}{3\pi^2(1+\bar{m}_k^2)^2}, \\
\label{Eq:largeNg1}
0&\equiv&\beta_{g_1}=-\bar{g}_{1,k}\nonumber\\
&&\hspace{-0.9cm}+\frac{2(\bar{g}_{1,k}^4+8\bar{g}^3_{1,k}\bar{g}_{2,k}+17\bar{g}_{1,k}^2\bar{g}_{2,k}^2+20\bar{g}_{1,k}\bar{g}_{2,k}^3+14\bar{g}_{2,k}^4)}{3\pi^2(\bar{g}_{1,k}+\bar{g}_{2,k})(\bar{g}_{1,k}+5\bar{g}_{2,k})(1+\bar{m}_k^2)^3},\nonumber\\ \\
\label{Eq:largeNg2}
0&\equiv&\beta_{g_2}=-\bar{g}_{2,k}+\frac{4\bar{g}_{2,k}^2(4\bar{g}_{2,k}-\bar{g}_{1,k})}{3\pi^2(\bar{g}_{1,k}+5\bar{g}_{2,k})(1+\bar{m}_k^2)^3}.
\eea
\end{subequations}
These algebraic equations can be easily solved numerically, and the found fixed points are shown in Table I. On top of the usual Gaussian and Wilson-Fisher [$O(2N_f^2 = \infty)$] fixed points, there are two additional ones with 1 ($C^{N_f=\infty}_1$) and 2 ($B^{N_f=\infty}_2$) relevant directions, respectively. Note that, subscripts refer to the number of relevant directions.

\begin{table*}
\centering
\vspace{0.2cm}
\label{Tab:N3}
  \begin{tabular}{ c | c | c | c | c | c | c | c }
  \hline
  ${\bf N_f}$ & FP & $\bar{m}^2$ & $\bar{g}_1$ & $\bar{g}_2$ & $\bar{a}$ & $\bar{b}$ & RD \\ \hline\hline
   ${\bf 3}$ & $O(2N_f^2)$ & $-0.31496$ & $0.43763$ & $0$ & $0$ & $0$ & $3$ $(2)$ \\ \hline
    $''$ & $\tilde{C}_2^3$ & $-0.38262$ & $0.59725$ & $-0.62042$ & $0$ & $0$ & $2$ $(1)$ \\ \hline
    $''$ & $A_4^3$ & $-0.01786$ & $0.091631$ & $-0.14148$ & $-0.11900$ & $0.39087$ & $4$ \\ \hline
    $''$ & ${A}_{1*}^3$ & $-0.41126$ & $0.73099$ & $-0.88199$ & $-0.46585$ & $-0.91131$ & $1*$ \\ 
  \hline  
  \end{tabular}
 \caption{Nontrivial fixed points for $N_f=3$. The number of relevant directions is shown in the last column. The numbers in the parentheses correspond to the relevant directions under the constraint that all directions corresponding to anomalous couplings disappear. Among the six fixed points with complex eigenvalues of the stability matrices, only the one with a single relevant direction is shown as ${A}_{1*}^3$.}
\end{table*}

\begin{table}[ht]
\centering
\vspace{0.2cm}
\label{Tab:N2}
  \begin{tabular}{ c | c | c | c | c | c }
  \hline
    ${\bf N_f}$ & FP & $\bar{m}^2$ & $\bar{g}_1$ & $\bar{g}_2$ & RD \\ \hline\hline
    ${\bf 2}$ & $O(2N_f^2)$ & $-0.27094$ & $0.85280$ & $0$ & $4$ $(3)$ \\ \hline
    $''$ &  $\tilde{C}^2_2$ & $-0.20599$ & $1.33367$ & $-1.88211$ & $2$ $(1)$\\ \hline
    $''$ & $\hat{C}^2_2$ & $-0.26318$ & $0.33093$ & $1.71728$ & $2$ $(1)$ \\
    \hline
  \end{tabular}
  \caption{Anomaly-free fixed points for $N_f=2$. The number of relevant directions is shown in the last column. The numbers in the parentheses correspond to the relevant directions under the constraint that all directions corresponding to anomalous couplings disappear.}
  \end{table}

\subsection{Cases of finite $N_f$}
As we move away from $N_f = \infty$, a new $\tilde{C}^{N_f}_1$ fixed point with one relevant direction branches from the $O(2N_f^2)$ fixed point. We show in Table II the values of the couplings in the fixed points for $N_f=50$, $N_f=20$, $N_f=10$, respectively. Once we hit $N_f=6$, the structure of the renormalization group flows changes, as the coefficient of the KMT determinant ceases to be (perturbatively) irrelevant and becomes marginal. Nevertheless, we see that even taking its effect into account, in all the found fixed points the corresponding coupling vanishes, therefore, it does not change the aforementioned structure.

\begin{table*}[t]
\centering
\vspace{0.2cm}
   \begin{tabular}{ c | c | c | c | c }
  \hline
    & $N_f=2$ & $N_f=3$ & $N_f=4$ & $N_f\geq 5$ \\ \hline \hline
    \makecell{$\epsilon$ expansion \\ ($\epsilon=1$)} & \hspace{0.1cm}2nd order*\hspace{0.1cm} & 1st order & 1st order & 1st order \\ \hline
    \makecell{FRG ($d=3$) \\ \{current study\}} 
    &  \makecell{\hspace{0.1cm}1st order (Case I) \hspace{0.1cm}  \\ \hspace{0.1cm}2nd order (Case II)\hspace{0.1cm} \\ \hspace{0.1cm}2nd order (Case III)\hspace{0.1cm}} 
    &  \makecell{\hspace{0.1cm}1st order (Case I) \hspace{0.1cm}  \\ \hspace{0.1cm}1st order (Case II)\hspace{0.1cm} \\ \hspace{0.1cm}2nd order (Case III)\hspace{0.1cm}} 
    &  \hspace{0.1cm}1st order \hspace{0.1cm}  
    & \hspace{0.1cm}2nd order\hspace{0.1cm}  \\ \hline
  \end{tabular}
  \caption{Comparison with axial anomaly [$SU(N_f)\times SU(N_f)$ model]. (*): In the $\epsilon$ expansion, for $N_f=2$, a second order transition only occurs, if the axial anomaly is strong enough at the critical point, leading to an $O(4)$ symmetric potential.}

\centering
\vspace{0.2cm}
   \begin{tabular}{ c | c | c | c | c }
  \hline
    & $N_f=2$ & $N_f=3$ & $N_f=4$ & $N_f\geq 5$ \\ \hline \hline
    \makecell{$\epsilon$ expansion \\ ($\epsilon=1$)} & 1st order & 1st order & 1st order & 1st order \\ \hline
    \makecell{FRG ($d=3$) \\ \{current study\}} & \hspace{0.1cm}2nd order\hspace{0.1cm} & \hspace{0.1cm} 2nd order\hspace{0.1cm} & \hspace{0.1cm} 2nd order\hspace{0.1cm} & \hspace{0.1cm} 2nd order\hspace{0.1cm} \\ \hline
  \end{tabular}
  \caption{Comparison without axial anomaly [$U(N_f)\times U(N_f)$ model].}
 
\end{table*}

The structure does change at $N_f=5$.  In this case, the KMT coupling is relevant, and we find only two fixed points on top of the trivial and the $O(2N_f^2)$ one, see Table III. $\tilde{C}_1^6\rightarrow \tilde{C}_1^5$ remains, but both $C^6_1$ and $B^6_2$ cease to exist, while a new fixed point, $A_3^5$ pops up with nonzero anomaly\footnote{All fixed points denoted by $A$ and its variations ($\tilde{A}$ or $\hat{A}$) have nonzero anomaly couplings.}, having three relevant directions. 

As for $N_f=4$, there are two anomalous couplings, one of them being marginal, while the other one relevant. The fixed point structure extends with another anomalous fixed point, $\tilde{A}_2^4$, see Table IV. Notice that, as the number of relevant operators increases, so does the number of relevant directions at each fixed point. As a result, there are no more fixed points with only one relevant direction. 

We find a similar conclusion for $N_f=3$, in accordance with the earlier study \cite{fejos22}, see Table V. As the number of relevant interactions become five (with four marginal ones), the flows are getting increasingly complicated, which is reflected in the number of the fixed points in the system. We again find an anomaly free $\tilde{C}_2^3$ fixed point with two relevant directions, and another one, $A_4^3$, with non-vanishing anomaly, but with four relevant directions. On top of these, we have six more fixed points with nonzero anomaly, whose stability matrices have complex eigenvalues; one of them, ${A}_{1*}^3$, has only one relevant direction, meaning that the real part of the eigenvalues is all but one positive. Whether the complex nature of the eigenvalues of ${A}_{1*}^3$ is only due to the LPA approximation is not clear, and we would not rule out the possibility that this fixed point indicates a second order chiral transition.

As already mentioned in the previous section, $N_f=2$ is the most complicated case. Since the coupling space is 12 dimensional (with six relevant and six marginal interactions), we could not find numerically all fixed points. We did obtain, however, all the anomaly-free fixed points as shown in Table VI. The $\tilde{C}^{N_f}$ fixed point continues to exist coming down from $N_f=3$ to $N_f=2$, retaining two relevant directions. On top of this, we see another fixed point with two relevant directions, that is $\hat{C}_2^2$. 

As for fixed points with nonvanishing anomalous couplings at $N_f=2$, we find an $O(4)$ symmetric one \cite{grahl13} associated to $\bar{m}^2=\infty$ and $|\bar{a}|=\infty$ with $\bar{m}^2+\bar{a}=$ being finite: It corresponds to the case where the $(\pi_0,\vec{s})$ multiplet becomes infinitely heavy, while $(s_0,\vec{\pi})$ has a finite mass. However, this fixed point is found to have the structure $\beta_{g_1}=\beta_{a_2}\neq \beta_{b_1}$ and $\beta_{\lambda_1}=\beta_{a_3}\neq \beta_{b_2}=\beta_{b_3}$, which are different from $\beta_{g_1}=\beta_{a_2}=\beta_{b_1}$ and $\beta_{\lambda_1}=\beta_{a_3}=\beta_{b_2}=\beta_{b_3}$, naively expected from the decoupling of $(\pi_0,\vec{s})$. It is not clear whether such a discrepancy is due to the LPA approximation. Note that this $O(4)$ fixed point has only one relevant direction, therefore, it could be responsible for a second order chiral transition at finite temperature.

In summary, for $5 \leq N_f < \infty$, there exists a collection of anomaly-free fixed points, $\tilde{C}_1^{N_f}$, which are continuously connected with one another, each having one relevant direction. At $N_f=4$, it continues to $\tilde{C}_2^{4}$, but now equipped with two relevant directions. The same is observed for $N_f=3$, and on top $\tilde{C}_2^{3}$, there also arises an anomalous fixed point with one relevant direction ($A_{1,*}^{3}$) but with complex eigenvalues of the stability matrix. For $N_f=2$, not only the anomaly-free fixed point, $\tilde{C}_2^{2}$ exists (again with two relevant directions) but also another one, $\hat{C}_2^{2}$ with the same properties, and also an anomalous $O(4)$ fixed point with one relevant direction.

We stress that, in case of the $U(N_f) \times U(N_f)$ model, none of the anomalous interactions exist in the free energy, therefore, the  locations of the anomaly-free fixed points, e.g., $\tilde{C}^{N_f=2,3,4}$, found in the $SU(N_f) \times SU(N_f)$ model, remain, while the number of their relevant directions decreases from two to one\footnote{In case of $N_f=2$, $\hat{C}_2^2\rightarrow \hat{C}_1^2$ also becomes a fixed point with only one relevant direction.}. That is to say, they may also correspond to finite temperature continuous phase transitions.

\subsection{Order of the transition}

At this point we consider three possibilities for the order of the finite temperature chiral transition in the $SU(N_f)\times SU(N_f)$ model.

\underline{{\it Case I (flavor continuity):}}  This corresponds to an assumption that the continuous family of anomaly free fixed points, $\tilde{C}^{N_f}$, dictates the chiral phase transition at finite temperature. It implies that the chiral transition is of second order for $N_f \ge 5$ [this statement is irrespective of the thermal fate of the $U_{\rm A}(1)$ anomaly], while for $N_f=2,3,4$, the transition is of first order. We note that if the anomaly disappears at the critical point [i.e., the $U(N_f)\times U(N_f)$ model applies], then the flavor continuity assumption yields the transition to be of second order even for $N_f=2,3,4$.

\underline{{\it Case II:}}  This corresponds to an assumption that the chiral transition is governed by the $\tilde{C}^{N_f}$ fixed points for $N_f \geq 3$, while for $N_f=2$ it is dictated by the anomalous O(4) fixed point. This case leads to the conclusion that the chiral transition is of second order for $N_f \geq 5$, first order for $N_f=3, 4$ and second order for $N_f=2$.

\underline{{\it Case III:}} This corresponds to an assumption that the chiral transition is governed by the $\tilde{C}^{N_f}$ fixed points for $N_f \geq 4$, while for $N_f=3$ and $N_f=2$ it is dictated by the anomalous $A_{1*}^3$ and O(4) fixed points, respectively. In such a case the chiral transition is of second order for $N_f \ge 5$, first order for $N_f=4$, and second order for $N_f=2, 3$.

In Table VII, we summarize the order of the chiral transition on the basis of the $\epsilon$ expansion and the present FRG study, for the $SU(N_f) \times SU(N_f)$ model. A similar summary for the $U(N_f) \times U(N_f)$ model can be found in Table VIII.

We note here that, although the exact recovery of $U_{\rm A}(1)$ symmetry is unlikely at any finite temperature \cite{laine04}, according to studies on the eigenvalue spectrum of the Dirac operator \cite{aoki12,kanazawa14,giordano24}, it cannot be ruled out that the anomaly becomes reasonably small at the critical point. In such a case, as the anomaly coefficient may control the strength of the transition\footnote{Similarly as an external magnetic field can control 
the strength of the first order transition in a ferromagnetic system.} (see the flavor continuity assumption above, for $N_f=2,3,4$), it being small enough might cause difficulties in distinguishing a
weak first order transition from a second order one. Turning this argument upside down, by constraining the strength of the transition, indirect conclusions could be drawn on the thermal behavior of the $U_{\rm A}(1)$ anomaly.

\begin{figure*}
\begin{center}
\raisebox{0.05cm}{
\includegraphics[bb = 0 320 495 570,scale=0.4,angle=270]{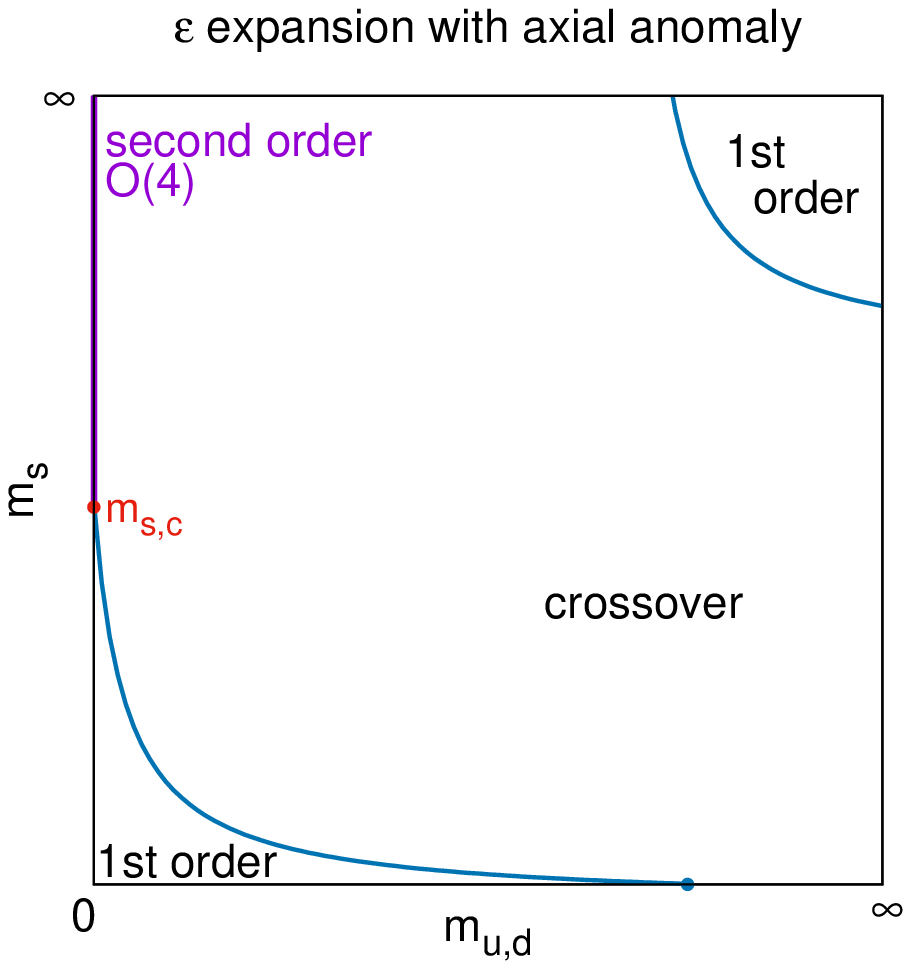}}
\includegraphics[bb = 5 80 495 410,scale=0.4,angle=270]{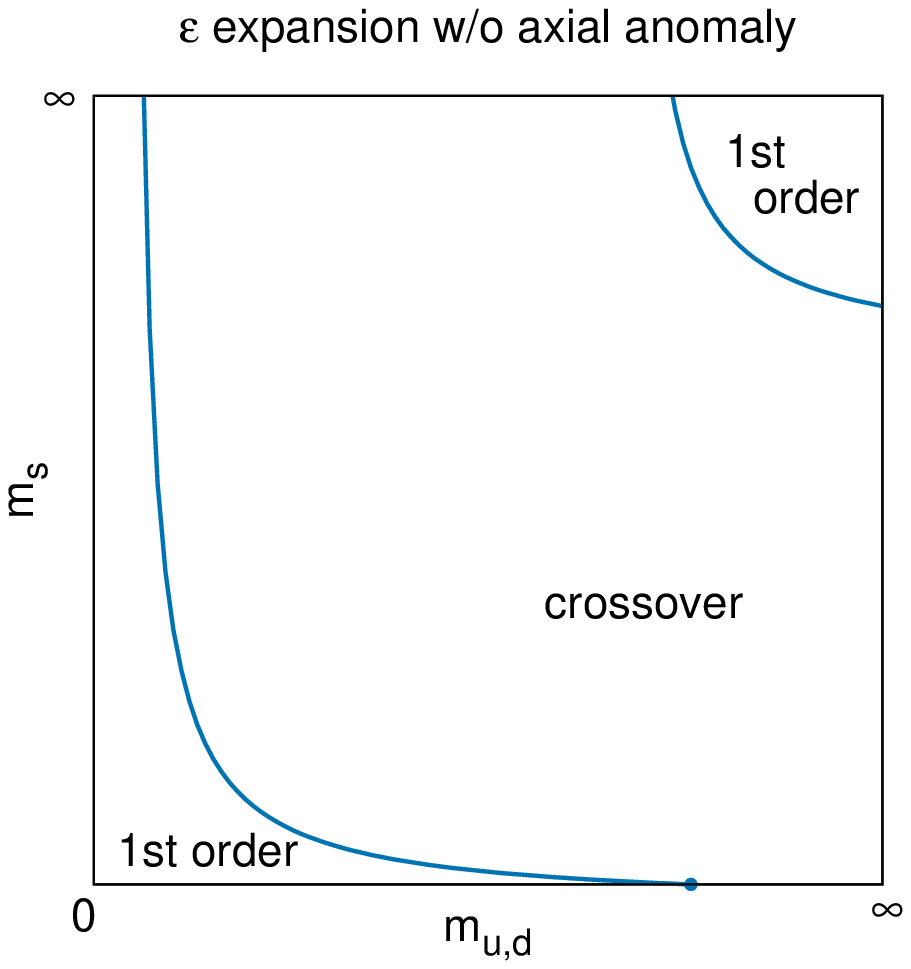}
\vspace{0.5cm} \caption{Columbia plot of the $\epsilon$ expansion. Left panel: 
the $U_{\rm A}(1)$ anomaly does not vanish at $T_c$, leading to a second order transition for $N_f=2$ (top left corner). Right panel: the $U_{\rm A}(1)$ anomaly vanishes exactly at $T_c$, leading to first order transitions for both $N_f=2,3$.}
\label{fig1}
\end{center}
\end{figure*}

\begin{figure*}
\begin{center}
\raisebox{0.05cm}{
\includegraphics[bb = 0 320 495 570,scale=0.4,angle=270]{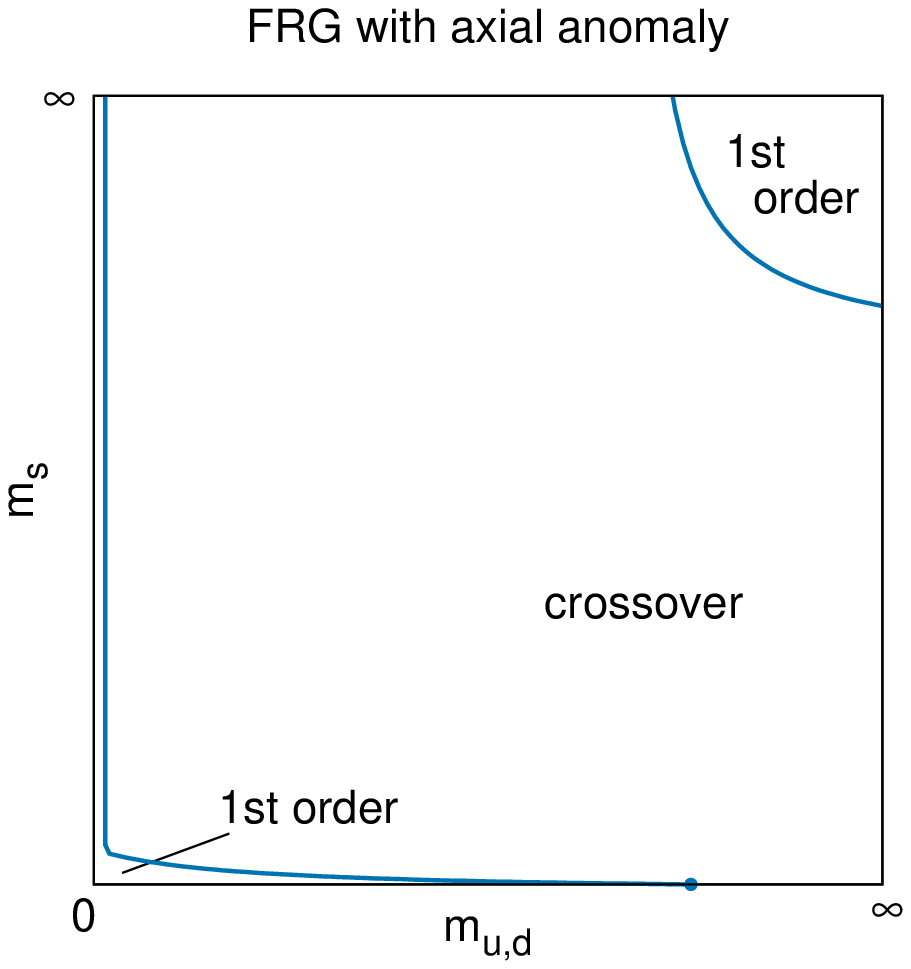}}
\includegraphics[bb = 5 80 495 410,scale=0.4,angle=270]{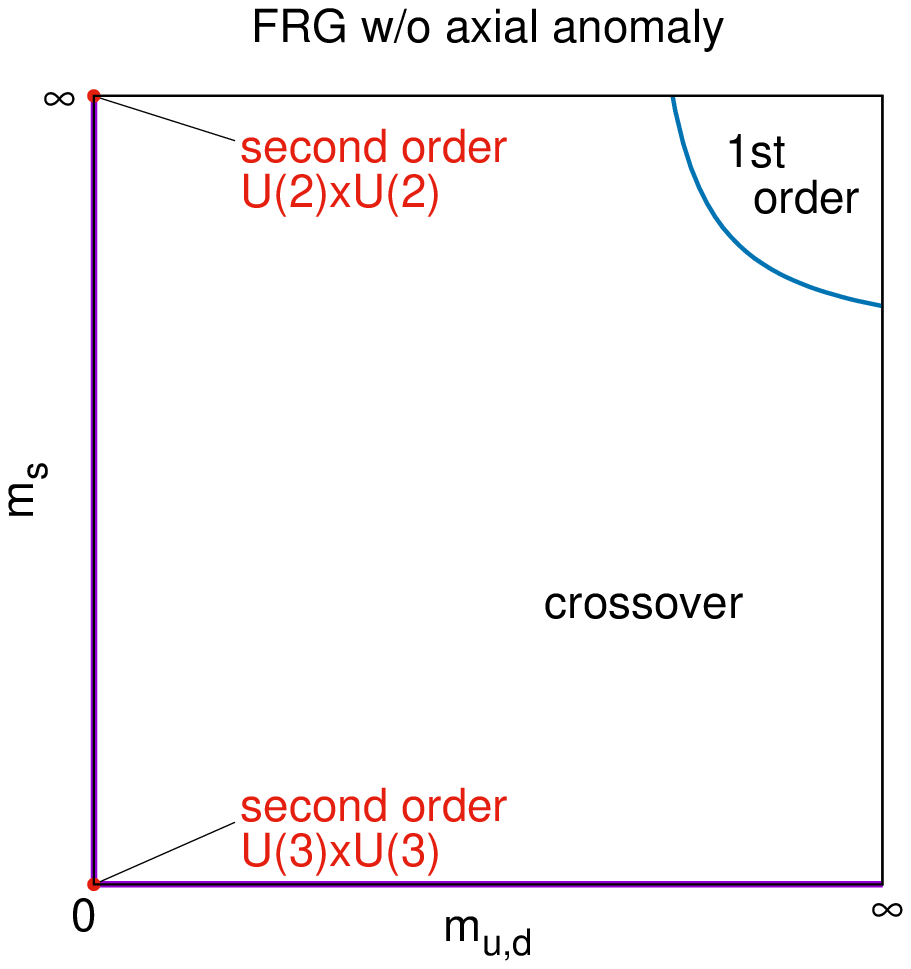}
\vspace{0.5cm} \caption{Columbia plot of the FRG method directly at $d=3$, assuming that the chiral transition is governed by the $\tilde{C}^{N_f}$ class of "flavor continuous" fixed points (case I in the text). Left panel: the $U_{\rm A}(1)$ anomaly does not vanish but is small at $T_c$, yielding the first order region very narrow. Right panel: the $U_{\rm A}(1)$ anomaly vanishes exactly at $T_c$, leading to second order transitions for both $N_f=2,3$ (top-left, bottom-left corners).}
\label{fig2}
\end{center}
\end{figure*}

\subsection{Relation to the $\epsilon$ expansion}

The $\epsilon$ expansion predicts that due to the absence of an IR stable fixed point, the chiral transition has to be of fluctuation induced first order for $N_f \geq 3$. This statement is irrespective of the thermal behavior of the $U_{\rm A}(1)$ anomaly\footnote{Note that, at $N_f=2$, for large anomaly the $\epsilon$ expansion also predicts the existence of an IR stable $O(4)$ fixed point.}, thus one may be curious of why our present study provides a different scenario, especially in light of the triumph of the $\epsilon$ expansion in $O(N)$-like models. 

First, we mention that had we dropped in our $d=3$ FRG approach all couplings that are not present in the original Pisarski-Wilczek study \cite{pisarski84}, we would have obtained exactly the same conclusion. Second, instead of working in $d=3$, the flow equation (\ref{Eq:flow}) can also be evaluated in $d=4-\epsilon$ dimensions, which leads to the $\beta$ functions obtained in \cite{pisarski84} at the leading order of a small $\epsilon$ expansion, as already shown in \cite{fejos14}.

We believe that the main reason why the $d=3$ FRG approach provides such different results compared to the $\epsilon$ expansion evaluated at $\epsilon=1$ is not that $\epsilon$ is ``not small enough'', but rather the operator structure of the free energy. Close to $d=4$ the number of operators is restricted, even if one adds gradually more marginal and irrelevant interactions, they can never produce any difference at the leading order of the $\epsilon$ expansion. We also think that the reason why the $\epsilon$ expansion works really well for the $O(N)$ model is that there is only one term difference between the ultraviolet free energy functional for $d=3$ and close to $d=4$, which cannot alter the fixed point structure. The $SU(N_f)\times SU(N_f)$ symmetry allows for a much larger set of invariants as we lower dimensionality, which creates space for new fixed points, yielding a much richer structure that can lead to critical behavior. We believe that the $\epsilon$ expansion can only work, if the number of invariant operators allowed by symmetry does not differ much for $d=3$ and $d=4$.

\section{Concluding remarks}

In this paper, we investigated the order of the chiral transition in the $SU(N_f)\times SU(N_f)$ model for a general flavor number, $N_f$. Renormalization group flows of the Ginzburg-Landau potential were calculated using the FRG method with the inclusion of all relevant and marginal interactions at $d=3$, i.e., up to ${\cal O}(\phi^6)$. The main difference compared to the $\epsilon$ expansion in $d=4-\epsilon$ dimensions is that the number of relevant and marginal interactions are considerably larger in the $d=3$ case.

We found new anomaly-free fixed points spanned throughout the entire range of $N_f$, which we call ``flavor continuous" fixed points, denoted by $\tilde{C}^{N_f}$. By conjecturing that the finite temperature chiral phase transition is governed by the $\tilde{C}^{N_f}$ fixed points, numerical analyses of the stability matrices around the fixed points show that the transition for $N_f \geq 5$ is of second order, irrespective of the realization of $U_{\rm A}(1)$ symmetry at the critical temperature. For $N_f = 2, 3, 4$, our results show that the transition is of first order, unless $U_{\rm A}(1)$ symmetry {\it exactly} recovers at the critical temperature \footnote{It may be natural to think that for odd $N_f$, the transition can only be of second order if the KMT coupling vanishes; otherwise no parity symmetric potential can be formed at the critical point \cite{pisarski24}. Our results indicate that fluctuations appear to be able to wash out the odd powered anomalous terms for $N_f \geq 5$, but it is not the case for $N_f=3$.}. (More general possibilities of the thermal chiral transition beyond the case of flavor continuity are summarized in Table VII.) We also argued that, in case that the anomaly is sufficiently weak around the critical temperature, it could be difficult to distinguish a first order transition from a second order one, and thus the critical quark mass ($m_{q,\rm{crit}}$) might be difficult to be extracted from numerical lattice QCD simulations.

In Figs. 1 and 2, we show schematic Columbia plots conjectured from 
the $\epsilon$ expansion and the flavor-continuity scenario of the FRG method, respectively. Left panels of these figures are for the axial anomaly being present at the critical temperature, while right panels are for the axial anomaly vanishing at the critical temperature. A small $m_{q, \rm{crit}}$, suggested by recent lattice QCD simulations are also considered in the left panel of Fig. 2.

We should still be cautious in drawing final conclusions as there is plenty of room for improving the present study. We have completely neglected the wave function renormalization factor (and thus the anomalous dimension), which may play an important role in the fixed point structure. Since our truncation was based on relevance (and irrelevance) around the Gaussian fixed point, it may be appropriate to test the robustness of the results with respect to introducing even higher order (perturbatively irrelevant) terms. Furthermore, the FRG formalism, through the flow equation (\ref{Eq:flow}), also makes it possible to investigate global fixed point potentials nonperturbatively, i.e., without expanding the potential in terms of the field variables. It would certainly be very important to obtain the corresponding potentials numerically, presumably on a grid, and investigate their stability.

\section*{Acknowledgments} 

The authors are grateful for Owe Philipsen, Rob Pisarski, and Fabian Rennecke for their useful remarks on the manuscript. This research was supported by the Hungarian National Research, Development, and Innovation Fund under Project No. FK142594. G. F. was supported by the Hungarian State Eötvös Scholarship. T. H. was supported by the Japan Science and Technology Agency (JST) as part of Adopting Sustainable Partnerships for Innovative Research Ecosystem (ASPIRE), Grant No. JPMJAP2318.

\makeatletter
\@addtoreset{equation}{section}
\makeatother 

\renewcommand{\theequation}{A\arabic{equation}} 

\

\ 

\appendix
\section{$U(N)$ algebra}   

The Lie algebra of the $U(N)\equiv U(1)\times SU(N)$ group contains $N$ diagonal and $N(N-1)$ nondiagonal generators. All but one are traceless and we choose the normalization as $\Tr(T_aT_b)=\delta_{ab}/2$. The diagonal ones, including the zeroth generator corresponding to the $U(1)$ subgroup read as
\vspace{0.1cm}
\bea
\!\!\!T_{0}&=&{\frac{1}{\sqrt{2N}}}
\left( \begin{array}{cccc}
1 & & & \\
& 1 & & \\
& & ... & \\
& & & 1 \\
\end{array} \right), \nonumber\\
\!\!\!T_{1}&=&\frac12
\left( \begin{array}{cccc}
1 & & & \\
& -1 & & \\
& & 0 & \\
& & & ... \\
\end{array} \right),\nonumber\\
\!\!\!T_{2}&=&\frac{1}{2\sqrt3}
\left( \begin{array}{cccc}
1 & & & \\
& 1 & & \\
& & -2 & \\
& & & ... \\
\end{array} \right),\nonumber\\
&...&\nonumber
\eea
\bea
\!\!\!T_{N-1}&=&\frac{1}{\sqrt{2N(N-1)}}
\left( \begin{array}{cccc}
1 & & & \\
& 1 & & \\
& & ... & \\
& & & -(N-1) \\
\end{array} \right).
\eea
We note that throughout the text we follow the notation $T_{N-1}\equiv T_L$, referring to the ``longest'' diagonal generator. The nondiagonal generators form two groups, as generalizations of the $\sigma_x$ and $\sigma_y$ Pauli matrices in the following sense:
\begin{subequations}
\label{app-nondiag}
\bea
\big(T_x^{(jk)}\big)_{ab}&=&\frac12(\delta_{ak}\delta_{bj}+\delta_{aj}\delta_{bk}), \\
\big(T_y^{(jk)}\big)_{ab}&=&\frac{i}{2}(\delta_{ak}\delta_{bj}-\delta_{aj}\delta_{bk}),
\eea
\end{subequations}
We have a number of $N(N-1)/2$ for each the $x$ and $y$ type generators. The $d_{abc}$ symmetric, and the $f_{abc}$ antisymmetric structure constants are defined through the products of generators:
\bea
T_aT_b=\frac12(d_{abc}+if_{abc})T_c.
\eea
The following identities turn out to be useful for calculating them:
\begin{subequations}
\bea
\label{app-struct}
f_{abc}&=&-2i\Tr\big[[T_a,T_b]T_c\big], \\
d_{abc}&=&2\Tr \big[\{T_a,T_b\}T_c\big], 
\eea
\end{subequations}
where $[\hspace{0.15cm},\hspace{0.15cm}]$ and $\{\hspace{0.15cm},\hspace{0.15cm}\}$ refer to the commutator and anticommutator, respectively. Alternatively, we can also deduce
\begin{subequations}
\bea
f_{abc}&=&4\Im \Tr(T_aT_bT_c),\\
d_{abc}&=&4\Re \Tr(T_aT_bT_c). 
\eea
\end{subequations}
Since for the practical calculations we mostly use a background field of $\Phi = s_0T_0+s_LT_L$, the following structure constants will be needed:
\begin{widetext}
\bea
\label{Eq:structcon}
d_{0ij}&=&\sqrt{\frac{2}{N}}\delta_{ij}, \quad d_{LLL}=(2-N)\sqrt{\frac{2}{N(N-1)}}, \quad d_{Lij\neq 0,8}=\begin{cases} (2-N)\sqrt{\frac{1}{2N(N-1)}}\delta_{ij} , \hspace{0.1cm} \iff \hspace{0.1cm} i,j \in \{(x,jN),(y,jN)\} \\ \sqrt{\frac{2}{N(N-1)}}\delta_{ij}, \hspace{1.3cm} \elsee \end{cases},\nonumber\\
d_{iLL}&=&\sqrt{\frac{2}{N}}\delta_{i0}+d_{LLL}\delta_{iL}, \qquad f_{uLv}=\sqrt{\frac{N}{2(N-1)}}\left(\delta_{u,(y,jN)}\delta_{v,(x,jN)}-\delta_{u,(x,jN)}\delta_{v,(y,jN)}\right).
\eea
\end{widetext}

\renewcommand{\theequation}{B\arabic{equation}} 

\section{Mass matrices}   

In this appendix we discuss how to calculate the ${\cal V}_k''$ mass matrix for a general flavor number $N_f$, defined via (\ref{Eq:Vdiv}). In what follows, we only consider the chirally symmetric part, ${\cal V}_{\ch,k}$, in ${\cal V}_k$. Derivatives for the anomalous part, ${\cal V}_{\an,k}$, are only relevant for $N_f=2,3,4,5,6$, and in such cases all derivatives can be calculated in analogy with (\ref{App:der}),  using symbolic programming. We do not list the corresponding formulas separately. 

Denoting by $\phi_i$ either the scalar, $s_i$, or pseudoscalar, $\pi_i$ fields, we get
\bea
\label{App:der}
({\cal V}_k'')_{\phi_i \phi_j} &=& m^2 \frac{\partial^2 I_1}{\partial \phi_i \partial \phi_j} + 2g_1 \frac{\partial I_1}{\partial \phi_i}\frac{\partial I_1}{\partial \phi_j} + 2g_1I_1 \frac{\partial^2 I_1}{\partial \phi_i \phi_j} \nonumber\\
&+&g_2 \frac{\partial^2 I_2}{\partial \phi_i \partial \phi_j} + 6\lambda_1 I_1 \frac{\partial I_1}{\partial \phi_i}\frac{\partial I_1}{\partial \phi_j}+3\lambda_1 I_1^2 \frac{\partial^2 I_1}{\partial \phi_i \phi_j} \nonumber\\
&+& \lambda_2 \frac{\partial^2 I_1}{\partial \phi_i \partial \phi_j} I_2+ \lambda_2 \Big(\frac{\partial I_1}{\partial \phi_i}\frac{\partial I_2}{\partial \phi_j} + \frac{\partial I_1}{\partial \phi_j}\frac{\partial I_2}{\partial \phi_i}\Big) \nonumber\\
&+&\lambda_2 I_1 \frac{\partial^2 I_2}{\partial \phi_i \partial \phi_j} + g_3 \frac{\partial^2 I_3}{\partial \phi_i \partial \phi_j}.
\eea
For the derivatives, we have
\bea
\frac{\partial I_1}{\partial \phi_i} &=& \phi_i, \quad \frac{\partial^2 I_1}{\partial \phi_i \partial \phi_j} = \delta_{ij},
\eea
and then by reintroducing the invariants $\tilde{I}_2$ and $\tilde{I}_3$ [see (\ref{Eq:inv})], so that $I_2 = \tilde{I}_2 - \frac{1}{N_f}I_1^2$ and $I_3 = \tilde{I}_3 - \frac{3}{N_f}I_1 \tilde{I}_2 + \frac{2}{N_f^2}I_1^3$, we arrive at
\begin{subequations}
\bea
\frac{\partial I_2}{\partial \phi_i} &=& \frac{\partial \tilde{I}_2}{\partial \phi_i} - \frac{2}{N_f}I_1 \frac{\partial I_1}{\partial \phi_i},  \\
\frac{\partial^2 I_2}{\partial \phi_i \partial \phi_j}&=&\frac{\partial^2 \tilde{I}_2}{\partial \phi_i \partial \phi_j} - \frac{2}{N_f}\frac{\partial I_1}{\partial s_i} \frac{\partial I_1}{\partial s_j}-\frac{2}{N_f}\frac{\partial^2 I_1}{\partial s_i \partial s_j},  \\
\frac{\partial I_3}{\partial \phi_i} &=& \frac{\partial \tilde{I}_3}{\partial \phi_i} - \frac{3}{N_f}\frac{\partial I_1}{\partial \phi_i}\tilde{I}_2 - \frac{3}{N_f}I_1 \frac{\partial \tilde{I}_2}{\partial \phi_i}+ \frac{6}{N_f}I_1^2 \frac{\partial I_1}{\partial \phi_i}, \nonumber\\ \\
\frac{\partial^2 I_3}{\partial \phi_i \partial \phi_j} &=& \frac{\partial^2 \tilde{I}_3}{\partial \phi_i \partial \phi_j} - \frac{3}{N_f}I_1\frac{\partial^2 \tilde{I}_2}{\partial \phi_i \partial \phi_j} -\frac{3}{N_f}\frac{\partial^2 I_1}{\partial \phi_i \partial \phi_j} \tilde{I}_2\nonumber\\
&-& \frac{3}{N_f}\Big(\frac{\partial I_1}{\partial \phi_i}\frac{\partial \tilde{I}_2}{\partial \phi_j}+\frac{\partial I_1}{\partial \phi_j}\frac{\partial \tilde{I}_2}{\partial \phi_i}\Big)+\frac{12}{N_f^2}I_1\frac{\partial I_1}{\partial \phi_i}\frac{\partial I_1}{\partial \phi_j}\nonumber\\
&+&\frac{6}{N_f^2}I_1^2\frac{\partial^2I_1}{\partial \phi_i \partial \phi_j}.
\eea
\end{subequations}
Then, after introducing the shorthand notation of $A_i = \frac{1}{2}(s_a-i\pi_a)(s_b+i\pi_b)(d_{abi}+if_{abi})$, we have
\bea
\tilde{I}_2 = \frac12 A_kA_k, \quad \tilde{I}_3 = \frac14 A_kA_lA_m d_{klm},
\eea
which leads to
\begin{subequations}
\bea
\frac{\partial \tilde{I}_2}{\partial \phi_i} &=& A_k\frac{\partial A_k}{\partial \phi_i}, \quad \frac{\partial^2 \tilde{I}_2}{\partial \phi_i\partial \phi_j} = A_k\frac{\partial^2 A_k}{\partial \phi_i\partial \phi_j} + \frac{\partial A_k}{\partial \phi_i}\frac{\partial A_k}{\partial \phi_j},\nonumber \\ \\
\frac{\partial \tilde{I}_3}{\partial \phi_i} &=&\frac34 \frac{\partial A_k}{\partial \phi_i} A_l A_m d_{klm}, \\
\frac{\partial^2 \tilde{I}_3}{\partial \phi_i\partial \phi_j} &=& \frac34 \frac{\partial^2 A_k}{\partial \phi_i\partial \phi_j} A_l A_m d_{klm} + \frac32 \frac{\partial A_k}{\partial \phi_i} \frac{\partial A_l}{\partial \phi_j}A_m d_{klm}. \nonumber\\
\eea
\end{subequations}
All the derivatives of ${\cal V}_k$ can now be expressed in terms of derivatives of $A_i$, which are found to be
\begin{subequations}
\bea
\!\!\!\frac{\partial A_m}{\partial s_j}&=&s^a d_{ajm}+\pi^a f_{ajm}, \quad \frac{\partial A_m}{\partial s_j}=\pi^a d_{ajm}-s^a f_{ajm},\nonumber\\ \\
&&\frac{\partial^2 A_m}{\partial s_i \partial s_j}=d_{ijm}, \quad \frac{\partial^2 A_m}{\partial \pi_i \partial \pi_j}=d_{ijm}.
\label{Eq:App}
\eea
\end{subequations}
As explained in the main text, for $N_f>6$ we use the background field of $\Phi = s_0T_0+s_LT_L$. Therefore, substituting $s_i = s_0\delta_{i0}+s_L\delta_{iL}$ and $\pi_i=0$, we get
\begin{subequations}
\bea
\!\!\!\!\!\!\!\!\!\!\!\!\!\!\!\!A_i &=& \frac{s_0^2+s_L^2}{\sqrt{2N_f}}\delta_{i0}\nonumber\\
\!\!\!\!\!\!\!\!\!\!\!\!\!\!\!\!&&+\Big(\frac{2-N_f}{\sqrt{2N_f(N_f-1)}}s_L^2+\sqrt{\frac{2}{N_f}}s_0s_L\Big)\delta_{iL},\\
\!\!\!\!\!\!\!\!\!\!\!\!\!\!\!\!\frac{\partial A_m}{\partial s_i}&=&s_0\sqrt{\frac{2}{N_f}}\delta_{mi}+s_L d_{Lmi},\quad \frac{\partial A_m}{\partial \pi_i}=s_Lf_{iLm}.
\eea
\end{subequations}
Note that the second derivatives are background independent, see (\ref{Eq:App}). Using $A_i$ and its derivatives we can now build up the ${\cal V}_k''$ matrix in the $\Phi = s_0T_0+s_LT_L$ background, which is the starting point of the evaluation of the Wetterich equation.

\end{document}